\documentclass{pasj00}
\def\gtsima{$\; \buildrel > \over \sim \;$}
\def\gsim{\lower.5ex\hbox{\gtsima}}
\def\ltsima{$\; \buildrel < \over \sim \;$}
\def\lsim{\lower.5ex\hbox{\ltsima}}
\begin{document}
\SetRunningHead{J.N.\ Winn et al.}
{Limits on H$\alpha$ Absorption by Transiting Planet HD~209458b}
\Received{2004/04/16}%{yyyy/mm/dd}
\Accepted{2004/05/06}%{yyyy/mm/dd}

\title{A Search for H$\alpha$ Absorption in the Exosphere\\
of the Transiting Extrasolar Planet HD~209458b}

\author{
Joshua N.\ \textsc{Winn}\altaffilmark{1}$^{*}$,
Yasushi \textsc{Suto}\altaffilmark{2},
Edwin L.\ \textsc{Turner}\altaffilmark{3},
Norio \textsc{Narita}\altaffilmark{2},\\
Brenda L.\ \textsc{Frye}\altaffilmark{3}$^{*}$,
Wako \textsc{Aoki}\altaffilmark{4}, 
Bun'ei \textsc{Sato}\altaffilmark{4,5}, and 
Toru \textsc{Yamada}\altaffilmark{4}
}

\bigskip

\altaffiltext{1}{Harvard-Smithsonian Center for Astrophysics, 60
Garden St.\ MS--51, Cambridge, MA 02138, USA}
\email{jwinn@cfa.harvard.edu}

\altaffiltext{2}{Department of Physics, School of Science, The
University of Tokyo, Tokyo 113-0033} \email{suto,
narita@utap.phys.s.u-tokyo.ac.jp}

\altaffiltext{3}{Princeton University Observatory, Peyton Hall,
Princeton, NJ 08544, USA} \email{bfrye, elt@astro.princeton.edu}

\altaffiltext{4}{National Astronomical Observatory of Japan, 2--21--1,
Mitaka, Tokyo 181--8588} \email{aoki.wako@nao.ac.jp,
bunei.sato@nao.ac.jp, yamada@optik.mtk.nao.ac.jp}

\altaffiltext{5}{Graduate School of Science and Technology, Kobe
University, 1-1 Rokkodai, Nada, Kobe 657-8501}


\maketitle

\bigskip

\begin{abstract}

\bigskip

There is evidence that the transiting planet HD~209458b has a large
exosphere of neutral hydrogen, based on a 15\% decrement in
Lyman-$\alpha$ flux that was observed by Vidal-Madjar et al.\ during
transits.  Here we report upper limits on H$\alpha$ absorption by the
exosphere.  The results are based on optical spectra of the parent
star obtained with the Subaru High Dispersion Spectrograph.
Comparison of the spectra taken inside and outside of transit reveals
no exospheric H$\alpha$ signal greater than 0.1\% within a 5.1~\AA\
band (chosen to have the same $\Delta\lambda/\lambda$ as the 15\%
Ly$\alpha$ absorption).  The corresponding limit on the column density
of $n=2$ neutral hydrogen is $N_2 \lsim 10^9$~cm$^{-2}$. This limit
constrains proposed models involving a hot ($\sim$10$^4$~K) and
hydrodynamically escaping exosphere.

\end{abstract}
\footnotetext[*]{National Science Foundation Astronomy \& Astrophysics
Postdoctoral Fellow}

\section{Introduction}

A milestone in extrasolar planet research was reached when Charbonneau
et al.\ (2000) and Henry et al.\ (2000) observed the photometric
signal of transits by the low-mass companion of HD~209458. The
companion was originally discovered by radial velocity measurements
(Mazeh et al.\ 2000), which specified only the orbital period (3.5
days), orbital eccentricity ($e<0.03$), and minimum mass of the
companion ($M\sin i = 0.69$~$M_{\rm Jup}$). The transit light curves
allowed the measurement of the companion's mass (by breaking the $\sin
i$ degeneracy) and radius (from the depth of the transit), providing
an unambiguous case of a planetary-mass companion to a main sequence
star, and a demonstration that the companion's density was that of a
gas giant planet.

Aside from this historic importance, the discovery made possible a
number of unique and important follow-up studies, based on more subtle
changes in the received starlight that should occur during transits.
In this paper we are concerned with changes due to the passage of a
small fraction of the starlight through the planetary atmosphere.
Selective absorption by atmospheric constituents causes the transit
depth to depend upon wavelength, an effect that forms the basis of
``transmission spectroscopy.'' Upper limits on various atmospheric
absorption features have been given by Bundy and Marcy (2000), Moutou
et al.\ (2001), Brown, Libbrecht, and Charbonneau (2002), and Moutou
et al.\ (2003). The first successful detection using this technique
was by Charbonneau et al.\ (2002), who observed a $(0.023\pm 0.006)$\%
increase in transit depth in the yellow light of the sodium resonance
doublet. A strong sodium signal had been predicted by Seager and
Sasselov (2000), and subsequently by Brown (2001) and Hubbard et al.\
(2001).

More recently, Vidal-Madjar et al.\ (2003) reported a remarkably large
transit depth of $(15\pm 4)$\% in the ultraviolet light of the
Lyman-$\alpha$ (Ly$\alpha$) transition of neutral hydrogen. To produce
such a large absorption, a spherical cloud of neutral hydrogen would
need a radius of $4.3R_{\rm Jup}$ or greater (depending on the optical
depth), as compared to the planetary radius of $1.3R_{\rm Jup}$
inferred from broad-band observations, and the Roche lobe radius of
$3.6R_{\rm Jup}$. Such a signal could be produced by an extended
envelope of hydrogen atoms that are evaporating from the planet's
upper atmosphere.  Vidal-Madjar et al.\ (2003) noted that the apparent
blueshift of the absorbing atoms (up to $-130$~km~s$^{-1}$) provided
additional evidence for atmospheric escape, and speculated that this
evaporation process accounts for the rarity of extrasolar planets with
orbital periods smaller than 3 days. Additional data obtained by
Vidal-Madjar et al.\ (2004) were consistent with the previous
Ly$\alpha$ result, and provided evidence (at the 2--3$\sigma$ level)
for exospheric absorption by oxygen (O~{\sc i}) and carbon (C~{\sc
ii}).

Further confirmation of the existence of this ``exosphere,'' and
characterization of its temperature, density, and composition are
clearly important goals.  Unfortunately, theoretical predictions for
the properties of the exosphere are not robust, making it difficult to
design strategies for observing the exosphere.  One approach is to
search for additional effects of hydrogen, which has already been
implicated by the Ly$\alpha$ measurement. If a significant fraction of
the hydrogen exists in the first excited state ($n=2$), it will
produce extra absorption in the Balmer lines of the stellar spectrum,
and in particular H$\alpha$ (6563~\AA).

Thus we were motivated to search for H$\alpha$ absorption from the
exosphere of HD~209458b.  A positive detection would confirm the
existence of the hydrogen exosphere and constrain its density and
temperature, with implications for the long-term evolution of ``hot
Jupiter'' extrasolar planets.  It would also represent a new and more
practical means of searching for exospheres of other extrasolar
planets, given the necessity of observing Ly$\alpha$ absorption above
the Earth's atmosphere, and the confusing effects of Ly$\alpha$
absorption by interstellar hydrogen and emission from the geocorona.
Previously, Bundy and Marcy (2000) placed upper limits of 3--4\% on
H$\beta$ and H$\gamma$ absorption in a 0.3~\AA\ band, although they
did not examine H$\alpha$ due to contamination of that region of the
spectrum by I$_2$ lines. Moutou et al.\ (2001) also searched for
exospheric absorption in optical spectra. They did not remark on
H$\alpha$ in particular, but set upper limits of $\approx$1\% for any
features of width 0.2~\AA, and noted that the limits could be improved
significantly with higher quality spectra taken in better atmospheric
conditions.

We have been obtaining optical echelle spectra of HD~209458 with the
Subaru Telescope, over a wide range of planetary orbital phases, in
order to search for reflected light from the planet.  Data taken on
the night of a planetary transit, described in section 2, are
well-suited for the H$\alpha$ search.  The procedure by which we
compared the spectra taken at different times is given in section 3,
and significantly improved upper limits on exospheric H$\alpha$
absorption are derived from the data. In section 4 we discuss the
physical interpretation and provide a brief summary.

\section{Observations and Data Reduction}

We observed HD~209458 on UT~2002~October~25 with the Subaru 8.2~m
telescope of the National Astronomical Observatory of Japan, on Mauna
Kea, Hawaii. The planet was predicted to transit its parent star in
the middle of that night, according to the ephemeris of Brown et al.\
(2001). We used the High Dispersion Spectrograph (HDS; Noguchi et al.\
2002) mounted at the $f/12.5$ Optical Nasmyth focus. The entrance slit
was $4\arcsec$ long and $0\farcs8$ wide, oriented at a constant
position angle. We used the standard Yb setup, in which the beam
passes through an order-blocking KV370 filter, is collimated by the
red-optimized mirror, dispersed by the echelle grating
(31.6~grooves~mm$^{-1}$, blaze angle $\timeform{71D.25}$),
cross-dispersed by the red-optimized grating (250~grooves~mm$^{-1}$,
blaze angle $\timeform{5D.00}$), and detected on two CCDs, each having
$4100\times 2048$ pixels of size 13.5~$\mu$m. The wavelength coverage
was 4100~\AA~$< \lambda <$~6800~\AA\ with resolution $R\approx 90000$.

We obtained 29 spectra of HD~209458, each with an exposure time of
500~seconds, through air masses ranging from 1.0 to 1.9.
Additionally, at both the start and end of this series of exposures,
we obtained a spectrum with an I$_2$ cell behind the slit (and a
narrower slit width of $0\farcs4$) in order to determine accurate
radial velocities for the star and verify the ephemeris. At the end of
the night, we obtained 7 spectra of the B5~Vn star HD~42545 in order
to estimate the telluric spectrum: any narrow absorption features in
the spectrum of this rapidly rotating star ($v\sin i =
245$~km~s$^{-1}$; Abt et al.\ 2002) are almost certainly telluric.
The weather conditions were excellent, and the seeing was
approximately $0\farcs6$.

We used standard IRAF\footnote{The Image Reduction and Analysis
Facility (IRAF) is distributed by the U.S.\ National Optical Astronomy
Observatories, which are operated by the Association of Universities
for Research in Astronomy, Inc., under cooperative agreement with the
National Science Foundation.} procedures to process the frames and
extract one-dimensional spectra. The frames were debiased, trimmed,
and multiplied by the gain. A median flat field was created from 20
dome-lamp exposures and normalized in two dimensions (along the
dispersion, to take out the lamp spectrum, and perpendicular to the
dispersion, to take out the aperture profile), and then used to
correct for pixel-to-pixel sensitivity variations. A two-dimensional
smooth polynomial surface was fitted to the inter-order scattered
light in each frame, and subtracted. The wavelength solution was
determined from thorium-argon arc exposures taken at the start and end
of each night.  A total of 54 orders were extracted, after fitting the
trace of each order of each spectrum separately, and summing the
counts across the dispersion with optimal weighting.  In the order
spanning H$\alpha$, the pixel scale was $\approx
0.02$~\AA~pixel$^{-1}$ and the full-width at half-maximum of arc lines
was approximately 4 pixels. The resulting one-dimensional spectra had
a typical signal-to-noise ratio of $\approx 300$ in each pixel.

\begin{figure*}[tbh]
\centering \FigureFile(80mm,80mm){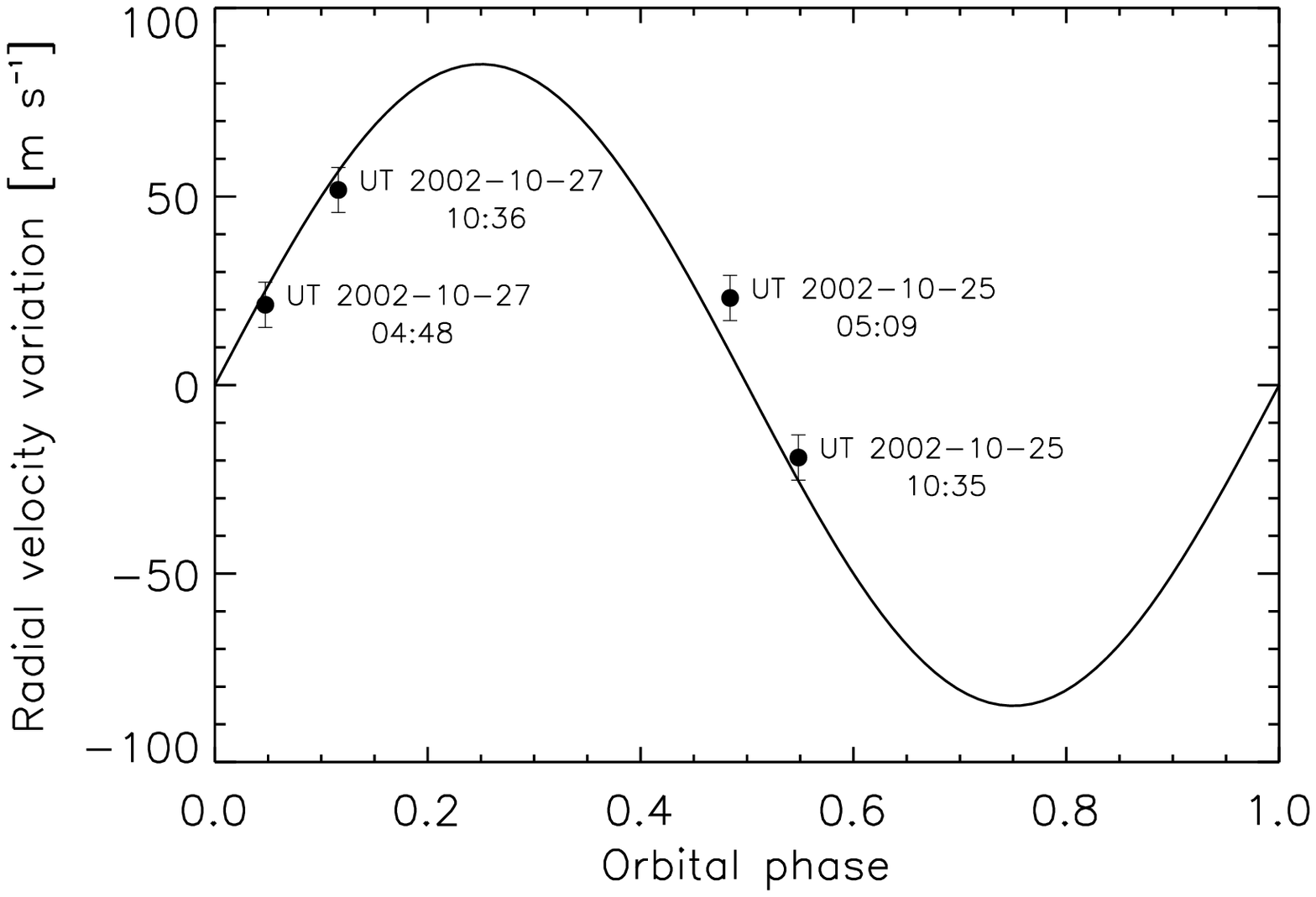} 
\caption{Subaru radial velocity measurements of HD~209458 on
2002~October~25 and 27.  The full radial velocity curve is also
plotted, using the orbital parameters determined by Naef et al.\
(2004). This confirms that the planetary transit took place between
the times of our two measurements on October~25, which bracketed the
time series of spectra presented in this paper.
\label{fig:rv_hd209458}}
\end{figure*}

Exposures taken with the I$_2$ cell were processed separately, using
the standard procedures described by \citet{Butler96} and
\citet{Sato02}. We determined the radial velocity based on the blue
CCD only.  Figure \ref{fig:rv_hd209458} shows our measurements, which
are uncertain by $\approx 6$~m~s$^{-1}$, as estimated from the
variance in results from 100 separate spectral segments in each
exposure. The two points from October~25, obtained at the start and
end of the HD~209458 observations on that night, bracket zero radial
velocity (orbital phase 0.5), confirming our expectation that the
transit took place between those times.

\section{Transmission Spectroscopy}

During the transit, the planet blocks a portion of the stellar
surface, causing the flux received from the star to decrease. Our goal
was not to measure the transit light curve, which would have required
observations of a comparison star to correct for flux variations due
to Earth's atmosphere. Rather, our goal was to measure differences
between the light curve observed in H$\alpha$ and at other
wavelengths, effectively using part of the parent star's spectrum as
the comparison signal. For exospheric effects to be detectable, they
must not only be large enough to overcome the Poisson noise in the
starlight, but they must also be distinguished from spectral
variations caused by instabilities in the Earth's atmosphere and the
spectrograph. In this section we describe the method we used to
attempt to make this distinction.

\subsection{Corrections for Instrumental Variations}

\begin{figure*}[tbh]
\centering \FigureFile(150mm,200mm){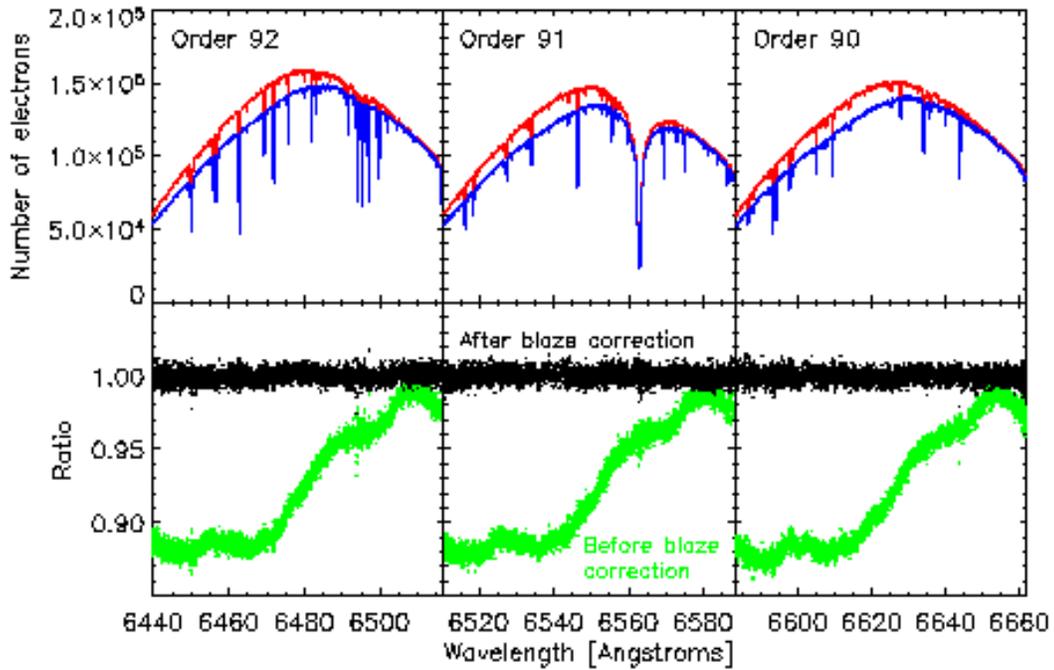}
\caption{ The top panels show sample spectra of HD~209458 taken
2.5~hours apart.  Each panel shows one order.  The bottom panels show
the ratio of the spectra, both before (green) and after (black) the
blaze function correction.\label{fig:differences} }
\end{figure*}

Figure~\ref{fig:differences} compares two representative spectra that
were taken 2.5~hours apart. The top three panels show three
consecutive orders bracketing H$\alpha$, the strong absorption feature
at 6563~\AA.  We refer to these spectra as $S_1(\lambda_{n,j})$ and
$S_2(\lambda_{n,j})$, where $\lambda_{n,j}$ is the $j$th wavelength
bin within the $n$th spectral order.  Note that we have not normalized
the spectra; the continua are peaked near the middle of each order due
to the blaze function of the spectrograph.

The green points in the bottom three panels of
figure~\ref{fig:differences} show the ratios $R(\lambda_{n,j}) =
S_1(\lambda_{n,j})/S_2(\lambda_{n,j})$. The mean of the ratio spectrum
is not unity, which is expected, because of variations in total flux
received in each exposure caused by the Earth's atmosphere. More
surprising are the wavelength variations in the ratio spectra: there
are $\pm 5$\% variations varying smoothly over $\approx 5$~\AA\
scales. We believe these variations are not an artifact of the
extraction procedure, because the same variations were obtained from
spectra that were extracted using different aperture sizes, and from
spectra that were extracted using custom-built procedures in the
Interactive Data Language (IDL) rather than IRAF.

The variations are surely instrumental, because the pattern is very
similar in each order: the ratio spectrum is a function of $j$ rather
than $\lambda$, and depends on $n$ to a much lesser degree. The ratio
spectra also exhibit a smooth time dependence. In particular, the
wavelength variations are greatest when one of the spectra involved in
the ratio was taken at the beginning of the night. This was when the
target was near zenith and the instrument rotator was moving most
quickly. The ratio spectra are flatter when the two spectra were taken
with nearly the same rotation angle. For these reasons, we hypothesize
that these apparent variations in the blaze function are caused by
flexure of the spectrograph. Similar variations have been seen by
other observers using Subaru/HDS (e.g., Nagao et al.\ 2003) and also
Keck/HIRES (e.g., Suzuki et al.\ 2003).

We corrected for this effect empirically by using the pattern observed
in the adjacent orders. To correct the $n$th order, we computed
$R(\lambda_{n+1,j})$ and $R(\lambda_{n-1,j})$, and then smoothed each
of these functions in $j$ with a boxcar size of 100 pixels (2~\AA),
giving $\tilde{R}(n+1,j)$ and $\tilde{R}(n-1,j)$. We then used linear
regression to determine the two numbers $c_{n+1}$ and $c_{n-1}$ that
minimized the sum of squared residuals between $S_1$ and
%%%%%%%%%%%%%%%%%%%%%%%%%%%%%%%%%%%%%%%%%%%%%%%%%%%%%%%%%%%%%%%%%%%%%
\small
\begin{equation}
S_{2{\rm b}}(\lambda_{n,j}) = \left[c_{n+1}\tilde{R}(n+1,j) +
                              c_{n-1}\tilde{R}(n-1,j) \right]
S_2(\lambda_{n,j}).
\end{equation}
\normalsize
%%%%%%%%%%%%%%%%%%%%%%%%%%%%%%%%%%%%%%%%%%%%%%%%%%%%%%%%%%%%%%%%%%%%%
For a given pair of spectra we typically found $c_{n+1}\approx
c_{n-1}\approx 0.5$. The resulting ``blaze corrected'' ratio spectrum,
$S_1/S_{2{\rm b}}$, is near unity with a standard deviation within 10\% of
that expected from the Poisson statistics (see black dots in
figure~\ref{fig:differences}).

Next, we corrected for small variations in the wavelength scale by
determining the two numbers $c_0$ and $\Delta\lambda$ that provided
the best match between $S_1$ and
%%%%%%%%%%%%%%%%%%%%%%%%%%%%%%%%%%%%%%%%%%%%%%%%%%%%%%%%%%%%%%%%%%%%%
\begin{equation}
S_{2{\rm m}}(\lambda_{n,j}) = c_0 S_{2{\rm b}}(\lambda_{n,j}) +
                  \Delta\lambda \frac{dS_{2{\rm b}}(\lambda)}{d\lambda},
\end{equation}
%%%%%%%%%%%%%%%%%%%%%%%%%%%%%%%%%%%%%%%%%%%%%%%%%%%%%%%%%%%%%%%%%%%%%
where the derivative was approximated numerically by 3-point
Lagrangian interpolation. The wavelength shifts were always smaller
than 0.005~\AA\ (0.25~pixel).  We refer to $S_{2{\rm m}}$ as the
``matched'' spectrum; it has been matched in blaze function and
wavelength scale to $S_1$.

\subsection{Calculation of Difference Spectra}

Having established this matching procedure to correct for instrumental
variations between any pair of spectra, we created a template spectrum
with a high signal-to-noise ratio in the following manner. A spectrum
from the middle of the night was chosen as a reference spectrum. All
the spectra taken through an air mass less than 1.4 were matched to
the reference spectrum. (Spectra taken through larger air masses were
significantly contaminated by time-variable telluric lines, as will be
seen shortly.) The preliminary template spectrum was defined as the
median of all the matched spectra. Then, the entire procedure was
repeated, using the preliminary template spectrum as the reference
spectrum, resulting in the final template spectrum $T(\lambda)$.
Finally, a time series of difference spectra was created, by
subtracting each spectrum from the suitably matched template.

\begin{figure*}[tbh]
\centering \FigureFile(175mm,195mm){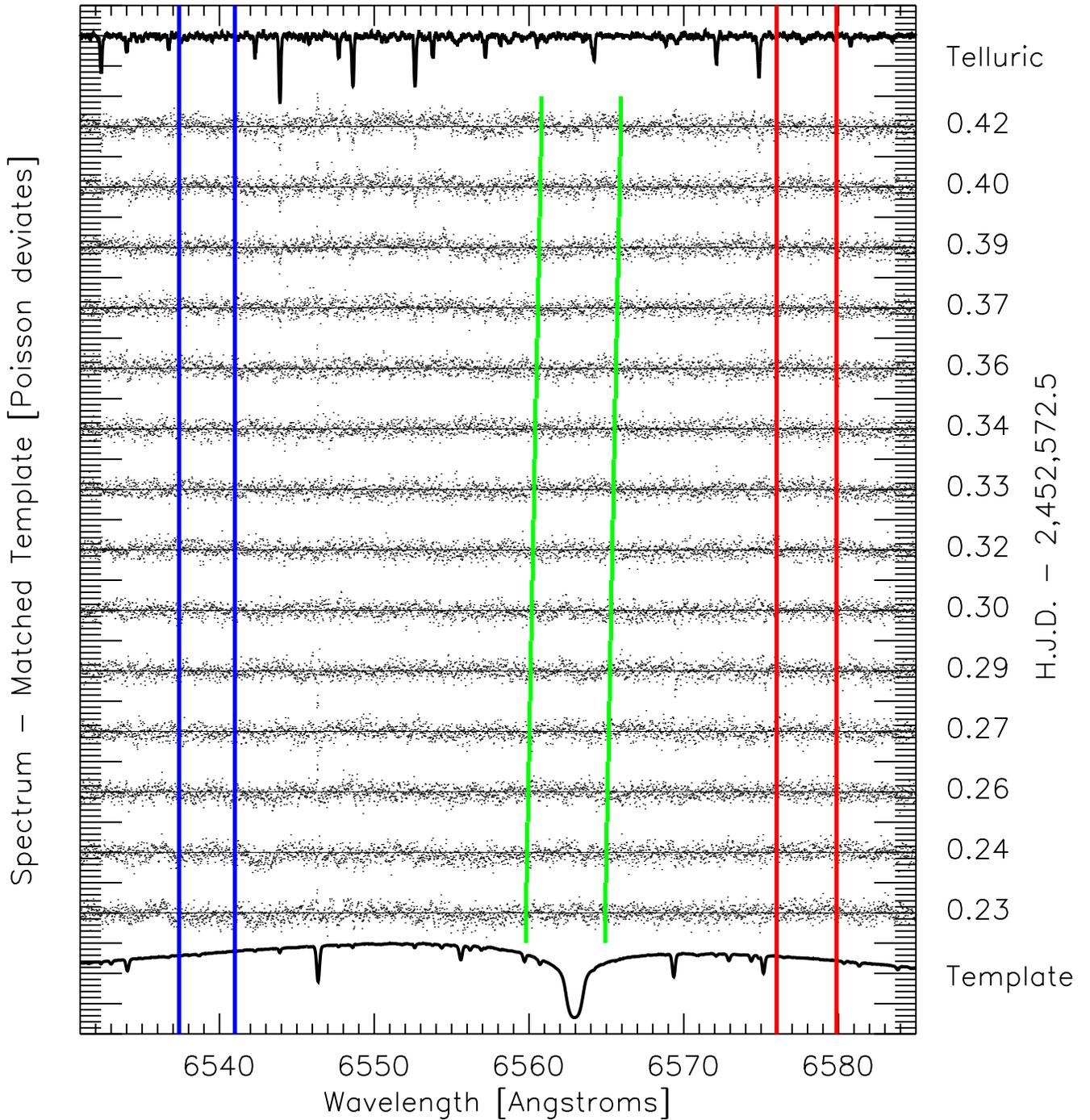}
\caption{ Differences between individual spectra of HD~209458 and the
matched template, expressed in units of Poisson deviates. The spacing
between minor tick marks represents one Poisson deviation. For
clarity, only 14 of the 29 spectra are shown. The template spectrum is
depicted at the bottom of the series, and the telluric spectrum is
depicted at the top. The right-hand axis shows the time each spectrum
was taken, in the same units as figure~\ref{fig:timeseries}. Colored
lines mark the edges of the three band passes described in the
text. \label{fig:residual_spectra} }
\end{figure*}

Some examples of the difference spectra are shown in
figure~\ref{fig:residual_spectra}, for a 55~\AA\ region in the
vicinity of H$\alpha$.  The difference spectra are plotted in units of
Poisson deviates,
%%%%%%%%%%%%%%%%%%%%%%%%%%%%%%%%%%%%%%%%%%%%%%%%%%%%%%%%%%%%%%%%%%%%%%
\begin{equation}
\frac{S(\lambda) - T_{\rm m}(\lambda)} {\sqrt{T_{\rm m}(\lambda)}},
\end{equation}
%%%%%%%%%%%%%%%%%%%%%%%%%%%%%%%%%%%%%%%%%%%%%%%%%%%%%%%%%%%%%%%%%%%%%%
with the space between minor tick marks indicating one Poisson
deviation.  Also plotted are the template spectrum (bottom) and the
estimated telluric spectrum (top).  To produce the telluric spectrum,
we created a template spectrum for HD~42545 in the same way as we did
for HD~209458.  Then we removed the rotationally broadened H$\alpha$
absorption line (which had a full-width at half-maximum of
$\sim$11~\AA) by fitting a smooth polynomial to the spectrum and
dividing by this polynomial.  We note that the telluric spectrum is
shown for visual reference only; we did not use it in any of the
calculations described below.

The difference spectra are nearly consistent with Poisson noise: they
have standard deviations near unity, as expected. However, a close
inspection of figure~\ref{fig:residual_spectra} shows the residuals
are not truly random. For spectra taken through large air masses (late
times), telluric lines are visible. In addition, there are excursions
of order 1--2$\sigma$ that are evident as ``ripples'' on scales of
2--5~\AA\ (100--250 pixels). For example, the spectrum at $t=0.24$
shows a sawtooth pattern in the residuals between 6535 and
6545~\AA. We do not know the cause of these ripples, but we are
confident that they are not due to the exosphere of the planet, for
two reasons: they are not specific to any particular absorption line
in the stellar spectrum; and they also appear in spectra taken on
October~27, when the planet was not transiting.

\subsection{Calculation of Difference Light Curves}

\begin{figure*}[th]
\centering \FigureFile(135mm,200mm){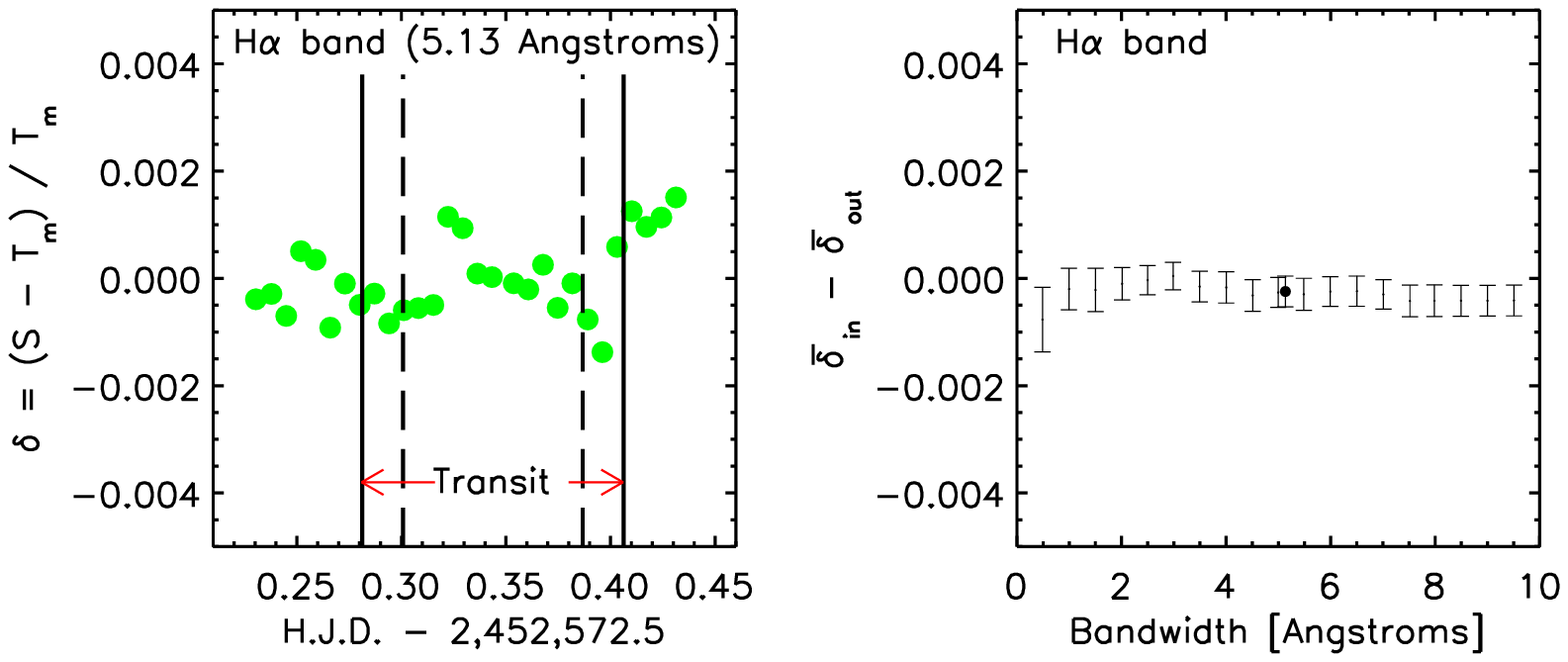}
\centering \FigureFile(135mm,200mm){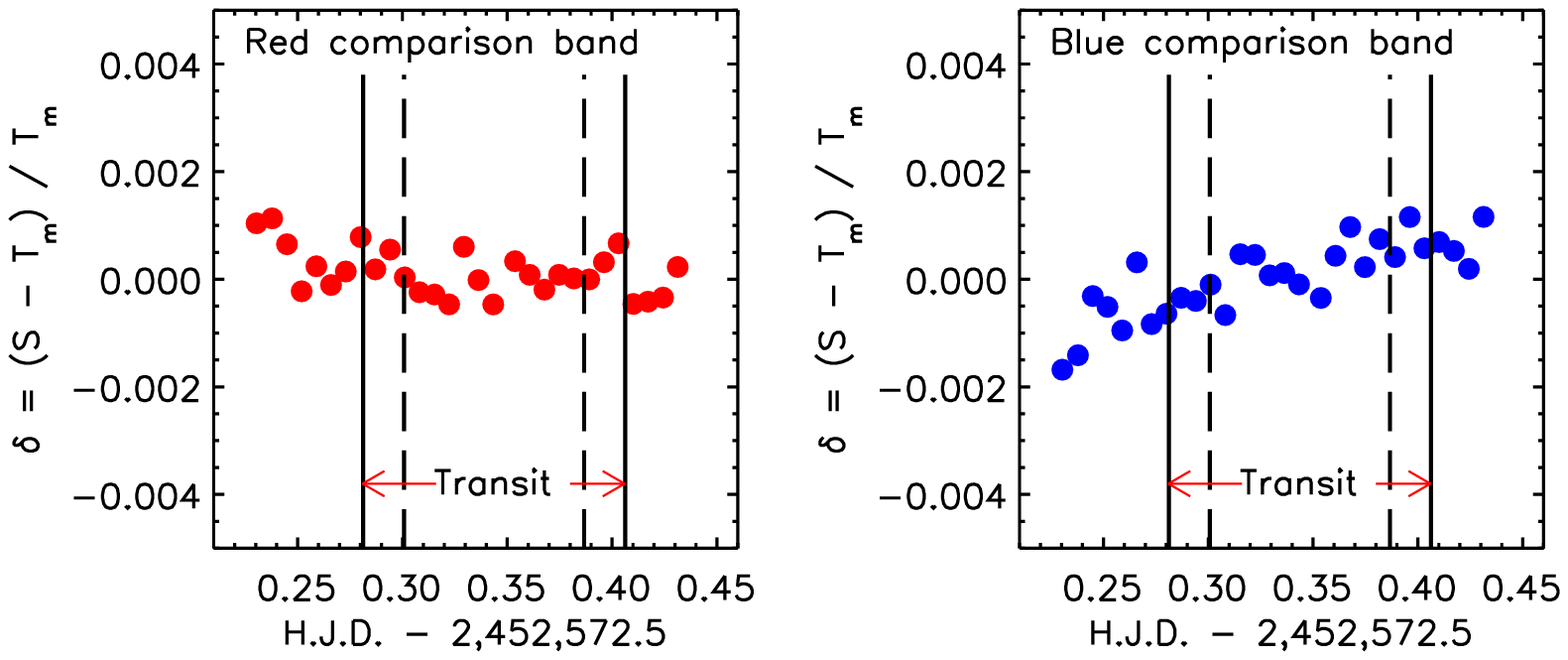}
\caption{ Comparison of the H$\alpha$ light curve and light curves in
adjacent bands. (Top left) The fractional difference in flux between
each spectrum $S$ and the corresponding matched template $T_{\rm m}$,
within the 5.1~\AA\ H$\alpha$ band shown in
figure~\ref{fig:differences}. Solid lines indicate the times of first
and fourth contact. Dashed lines indicate the times of second and
third contact. (Bottom left) Same, but for the red comparison
band. (Bottom right) Same, but for the blue comparison band. (Upper
right) The difference between in-transit and out-of-transit flux
deviations from the template, as a function of the width of the band
pass. The single filled point is for the 5.1~\AA\
band.\label{fig:timeseries} }
\end{figure*}

There are no obviously significant features in the difference spectra
at the position of H$\alpha$, implying that once the spectra have been
adjusted to correct for total flux and instrumental effects, there is
no significant difference between the H$\alpha$ light curve and those
of neighboring bands. To plot these light curves, we computed the
fractional difference in flux between the spectrum and the matched
template,
%%%%%%%%%%%%%%%%%%%%%%%%%%%%%%%%%%%%%%%%%%%%%%%%%%%%%%%%%%%%%%%%%%%%%%
\begin{equation}
\delta(t) =
   \frac{\displaystyle \int_{\lambda_1}^{\lambda_2} d\lambda 
\hspace{0.1in} \left[ S(\lambda, t) - T_{\rm m}(\lambda) \right] }
        {\displaystyle \int_{\lambda_1}^{\lambda_2} d\lambda 
\hspace{0.1in} T_{\rm m}(\lambda) }
\end{equation}
%%%%%%%%%%%%%%%%%%%%%%%%%%%%%%%%%%%%%%%%%%%%%%%%%%%%%%%%%%%%%%%%%%%%%%
for the 3 different wavelength bands indicated in
figure~\ref{fig:residual_spectra}: a band centered on H$\alpha$
(green), and two comparison bands (red and blue) in the adjacent
continua.  The width of the H$\alpha$ band (5.13~\AA) was chosen to
match the $\Delta\lambda / \lambda$ of the band over which
\citet{Vidal-Madjar03} reported Ly$\alpha$ absorption. The central
wavelength of the H$\alpha$ band was shifted to compensate for the
time-variable Doppler shift of the planetary orbit. This shift is
evident as a tilt in the green lines of Figure 3. The central
wavelengths of the comparison bands were chosen to avoid telluric
lines and strong stellar absorption lines, as much as possible. The
widths of the red (3.87~\AA) and blue (3.61~\AA) comparison bands were
chosen such that all three bands have the same total flux in the
template spectrum. Figure~\ref{fig:timeseries} shows the light curve
$\delta(t)$ for each of the three bands. Apart from a few outliers,
the variations in the H$\alpha$ band are similar in size and character
to the variations in the comparison bands. The red and blue light
curves exhibit small (and opposite) gradients which are probably
effects of increasing air mass.

To place a quantitative limit on excess H$\alpha$ absorption, we
calculated $\bar{\delta}_{\rm in}$, the mean value of $\delta(t)$ for
spectra taken during transit (between second and third contacts). As
an estimate of the uncertainty in $\bar{\delta}_{\rm in}$, we computed
the standard deviation of $\delta(t)$ for the 12 in-transit spectra
and divided by $\sqrt{12}$. Likewise, we computed $\bar{\delta}_{\rm
out}$ and its uncertainty from the spectra taken before first contact
or after fourth contact.\footnote{To compute the transit times, we
used an ephemeris defined by J.D.~(mid-transit) = 2451659.93675 (Brown
et al.\ 2001) and the most precise published period, $P=3.524739$~d
(Robichon and Arenou 2000).} Then we computed $\delta_{{\rm H}\alpha}
\equiv \bar{\delta}_{\rm in} - \bar{\delta}_{\rm out}$, which would be
negative if there were extra H$\alpha$ absorption during transit.  For
the 5.13~\AA\ band, the result is $\delta_{{\rm H}\alpha} = -0.00024
\pm 0.00029$. We investigated the dependence of this limit upon
bandwidth, by repeating this procedure for bands varying in width from
0.5~\AA\ to 10~\AA, centered on H$\alpha$. The results are shown in
the upper right panel of figure~\ref{fig:timeseries}.  Finally, we
tried blueshifting the 5.13~\AA\ H$\alpha$ band by $-130$~km~s$^{-1}$,
to match the possible blueshift in Ly$\alpha$ observed by Vidal-Madjar
et al.\ (2003).  The outcome was also a null result, $\delta_{{\rm
H}\alpha} = 0.00015 \pm 0.00021$.

\begin{figure*}[bh]
\centering \FigureFile(135mm,200mm){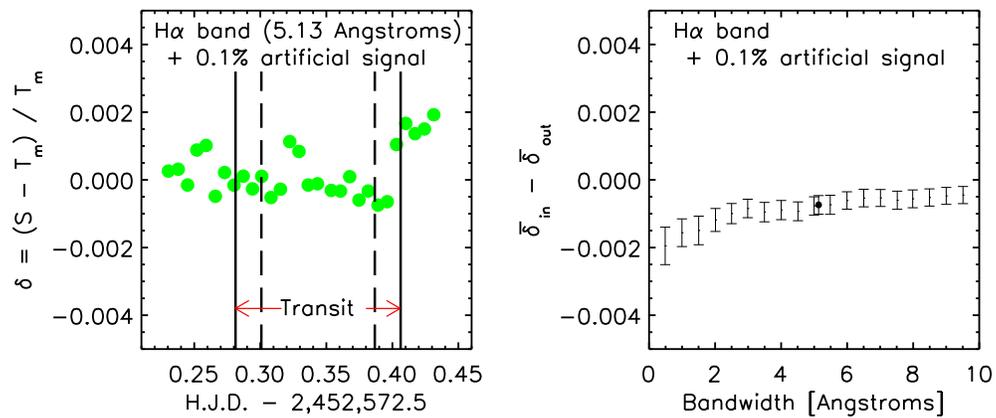}
\caption{ Same as the top two panels of figure~\ref{fig:timeseries},
except that an artificial absorption signal of strength 0.1\% has been
added to the in-transit spectra. From this test we concluded that a
0.1\% excess in H$\alpha$ absorption during transits would have been
detected.\label{fig:timeseries_fake} }
\end{figure*}

Although our spectrum-matching procedure was intended to correct only
for instrumental and telluric effects, it also acts to dilute any real
signal. To estimate the size of this effect we injected an artificial
signal into the data.  We multiplied the extracted spectra by
$(1-g_\lambda)$, where $g_\lambda$ is a Gaussian function centered on
H$\alpha$ with a full-width at half-maximum of 3~\AA\ (a fairly
arbitrary choice, since the true width of the absorbing feature is
unknown), representing a total flux decrement of 0.1\% in the 5.13~\AA\
band over which $\delta_{{\rm H}\alpha}$ was measured.  When the
preceding analysis was performed on the adulterated data, the result
was $\delta_{{\rm H}\alpha} = -0.00081\pm 0.00028$. This is greater by
0.00057 than the measurement with no artificial signal, verifying that
the artificial signal (although diluted) could be recovered in spite
of the noise. We conclude than an extra 0.1\% absorption during
transit would have been detected. Figure~\ref{fig:timeseries_fake}
shows the resulting light curve $\delta(t)$ and the ``measured''
absorption as a function of the width of the H$\alpha$ band.

\section{Discussion and summary}

Fundamentally, our null result corresponds to a maximum number of
neutral hydrogen atoms in the $n=2$ state that are present in the
exosphere of HD~209458b. Likewise, the detection of Ly$\alpha$
absorption by Vidal-Madjar et al.\ (2003) implies a minimum number of
neutral hydrogen atoms in the $n=1$ state.  Unfortunately it is
difficult to use these results to constrain interesting quantities
such as the exospheric temperature and density, because the physical
conditions in the exosphere are likely to be quite complex. The column
density of neutral hydrogen may vary widely across the surface of the
primary star due to evaporation, tidal forces, and radiation pressure
(see, e.g., Lecavelier des Etangs et al.\ 2004). Gas may be present at
a variety of temperatures and densities, and may be far from local
thermodynamic equilibrium (LTE). [Even in the denser atmosphere,
non-LTE effects may be required to explain the unexpectedly weak
detection of sodium by Charbonneau et al.\ (2002), as proposed by
Barman et al.\ (2002) and Fortney et al.\ (2003).]  Obviously the
limited empirical information available is not sufficient by itself to
determine a realistic physical exosphere models.

With these caveats, we can use a simple order-of-magnitude argument to
provide an upper limit on the column density of $n=2$ neutral hydrogen
that exosphere models will need to obey. In order to translate our
upper limit on additional H$\alpha$ absorption into a corresponding
upper limit on the H$\alpha$ equivalent width, we combined the
in-transit spectra to form $S_{\rm in}(\lambda)$, and we combined the
out-of-transit spectra to form $S_{\rm out}(\lambda)$.  Then the
equivalent width $W_\lambda$ of any transit absorption feature is
%%%%%%%%%%%%%%%%%%%%%%%%%%%%%%%%%%%%%%%%%%%%%%%%%%%%%%%%%%%%%%%%%%%%%%%%
\begin{eqnarray}
\label{eq:eqwid}
 W_\lambda \approx
 \displaystyle \int d\lambda \hspace{0.05in}
    \left[ \frac{ S_{\rm out}(\lambda) - S_{\rm in}(\lambda) }
                { S_{\rm out}(\lambda) }  \right].
\end{eqnarray}
%%%%%%%%%%%%%%%%%%%%%%%%%%%%%%%%%%%%%%%%%%%%%%%%%%%%%%%%%%%%%%%%%%%%%%%%
Performing the integral over the 5~\AA\ band described in the previous
section, we find $W_{{\rm H}\alpha} < 1.7$~m\AA.  If we further assume
that the fraction of the stellar surface covered by the exosphere is
$\Delta A / A \approx 0.15$ (as estimated from the Ly$\alpha$ result),
then the null result implies that H$\alpha$ absorption is weak, and
the corresponding limit on the column density of $n=2$ atoms ($N_2$)
is determined via
%%%%%%%%%%%%%%%%%%%%%%%%%%%%%%%%%%%%%%%%%%%%%%%%%%%%%%%%%%%%%%%%%%%%%%%%
\begin{eqnarray}
\label{eq:halpha-width}
   \left( \frac{\Delta A}{A} \right) W_{{\rm H}\alpha} =
   \frac{\pi e^2}{m c^2} f_{23} \lambda_{{\rm H}\alpha}^2 N_2,
\end{eqnarray}
%%%%%%%%%%%%%%%%%%%%%%%%%%%%%%%%%%%%%%%%%%%%%%%%%%%%%%%%%%%%%%%%%%%%%%%%
where $f_{23}\approx 0.641$ is the oscillator strength of the
$n=2\rightarrow 3$ transition.  The resulting limit is $N_2 < 1.0\times
10^9$~cm$^{-2}$.

In principle, $N_1$ could be estimated from the Ly$\alpha$ decrement
observed during transits, and the ratio $N_2/N_1$ could be related to
the H~{\sc i} excitation temperature. In reality, $N_1$ is highly
uncertain because the absorption is probably saturated, and the line
profile is unobservable due to interstellar absorption and geocoronal
emission.  However, any exosphere model that specifies $N_1$ is
constrained by our results to have a maximum excitation temperature
$T_{\rm ex}$, according to the relation
%%%%%%%%%%%%%%%%%%%%%%%%%%%%%%%%%%%%%%%%%%%%%%%%%%%%%%%%%%%%%%%
\begin{eqnarray}
\label{eq:n1n2ratio2}
\frac{N_2}{N_1} = \frac{g_2}{g_1} \exp \left(-\frac{E_2 - E_1}{kT_{\rm ex}} \right)
                = 4\hspace{0.03in} \exp\left(-\frac{10.2~{\rm eV}}{kT_{\rm ex}}\right).
\end{eqnarray}
%%%%%%%%%%%%%%%%%%%%%%%%%%%%%%%%%%%%%%%%%%%%%%%%%%%%%%%%%%%%%%%%
This constraint appears to be important in the context of recently
proposed models in which the exosphere is very hot and
hydrodynamically escaping, as argued below.

The temperature of the lower atmosphere of HD~209458b is likely to be
near the radiative effective temperature, $T_{\rm eff} \approx
10^3$~K, at which the $n=2$ population predicted from
equation~(\ref{eq:n1n2ratio2}) is utterly negligible. But, as argued
by Moutou et al.\ (2001), Lammer et al.\ (2003), and Lecavelier des
Etangs et al.\ (2004), the exosphere is likely to be significantly
hotter than $T_{\rm eff}$, just as the exosphere of Jupiter is
considerably hotter than its lower atmosphere ($\sim$10$^3$~K vs.\
150~K; see, e.g., Atreya 1986). Lammer et al.\ (2003) found that
close-in gas giant planets could have exospheric temperatures up to
$10^4$~K due to intense X-ray and extreme-ultraviolet (EUV)
irradiation. Depending on the density of neutral hydrogen and other
factors, such a hot exosphere could produce detectable H$\alpha$
absorption in violation of our constraint.

For example, Lecavelier des Etangs et al.\ (2004) presented 3 models
for the thermosphere and exosphere of HD~209458b. In each case they
specified the density profile of neutral hydrogen and computed the
size and shape of the exosphere as a function of kinetic temperature
$T_{\rm k}$, taking tidal distortion into account. For the specific
case of Model A, the density is $n_{\rm H I}=2\times 10^9$~cm$^{-3}$
at the base of the thermosphere, and decreases with radius according
to the barometric law. We computed the average column density $N_1$
within the annulus reaching from the thermobase to the Roche radius,
and then used equation~(\ref{eq:n1n2ratio2}) to translate our upper
limit on $N_2$ into an upper limit on the excitation temperature,
finding $T_{\rm ex} < 8000$~K. By comparison, Lecavelier des Etangs et
al.\ (2004) found that $T_{\rm k}> 8000$~K in order to produce an
escape rate of $10^{10}$~g~s$^{-1}$ [the lower limit inferred by
Vidal-Madjar et al.\ (2003)], and $T_{\rm k}> 11100$~K through a
simulation of EUV heating. Of course, one would not necessarily expect
$T_{\rm ex} \approx T_{\rm k}$ in the tenuous thermosphere and
exosphere, because of departures from LTE; in general, $T_{\rm ex}$ is
neither an upper bound nor a lower bound on $T_{\rm k}$. The $n=2$
state may be depopulated when collisions are infrequent, and likewise,
whatever process accelerates the gas to $\sim$100~km~s$^{-1}$ may
overpopulate the $n=2$ state via recombination cascades and Ly$\alpha$
resonant trapping. A full non-LTE exosphere model will need to
accommodate the resulting upper limit on the excitation temperature.

In summary, we have used high resolution and high signal-to-noise
ratio Subaru-HDS optical spectra to search for excess Balmer H$\alpha$
absorption during a transit of the extrasolar planet HD~209458b. No
excess absorption was detected. Our upper limit is two orders of
magnitude below the Ly$\alpha$ absorption reported by Vidal-Madjar et
al.\ (2003, 2004). It may be difficult to improve upon this limit with
ground-based instruments, given the difficulty of correcting for
telluric and instrumental spectral variations.  However, there should
be no obstacle to improving upon our limit with space-based
spectroscopy, and we are aware that such an effort is underway using
data from the Hubble Space Telescope (Charbonneau, D., private
communication). The current result, and any future refinements, will
be useful in the further development of models for the escape of the
hot exospheres of close-in gas giant planets.

\bigskip
%%%%%%%%%%%%%%%%%%%%%%%%%%%%%%%%%%%%%%%%%%%%%%%%%%%%%%%%%%%%%%%%%%%%%%

We are very grateful to Yutaka Abe, Dave Charbonneau, Bruce Draine,
Bob Kurucz, Avi Loeb, Paul Martini, Dimitar Sasselov, and Motohide
Tamura, for helpful consultations. This work is based on data from the
Subaru Telescope, which is operated by the National Astronomical
Observatory of Japan.  We wish to recognize and acknowledge the very
significant cultural role and reverence that the summit of Mauna Kea
has always had within the indigenous Hawaiian community.  We are most
fortunate to have the opportunity to conduct observations from this
mountain.

\bigskip
%%%%%%%%%%%%%%%%%%%%%%%%%%%%%%%%%%%%%%%%%%%%%%%%%%%%%%%%%%%%%%%%%%%%%%

%%%%%%%%%%%%%%%%%%%%%%%%%%%%%%%%%%%%%%%%%%%%%%%%%%%%%%%%%%%%%%%%%%%%%%
%%%%%%%%%%%%%%%%%%%%%%%%%%%%%%%%%%%%%%%%%%%%%%%%%%%%%%%%%%%%%%%%%%%%%%

%%%%%%%%%%%%%%%%%%%%%%%%%%%%%%%%%%%%%%%%%%%%%%%%%%%%%%%%%%%%%%%%%%%%%%
%%%%%%%%%%%%%%%%%%%%%%%%%%%%%%%%%%%%%%%%%%%%%%%%%%%%%%%%%%%%%%%%%%%%%%

%%%%%%%%%%%%%%%%%%%%%%%%%%%%%%%%%%%%%%%%%%%%%%%%%%%%%%%%%%%%%%%%%%%%%%
%%%%%%%%%%%%%%%%%%%%%%%%%%%%%%%%%%%%%%%%%%%%%%%%%%%%%%%%%%%%%%%%%%%%%%

%%%%%%%%%%%%%%%%%%%%%%%%%%%%%%%%%%%%%%%%%%%%%%%%%%%%%%%%%%%%%%%%%%%%%%
%%%%%%%%%%%%%%%%%%%%%%%%%%%%%%%%%%%%%%%%%%%%%%%%%%%%%%%%%%%%%%%%%%%%%%

%%%%%%%%%%%%%%%%%%%%%%%%%%%%%%%%%%%%%%%%%%%%%%%%%%%%%%%%%%%%%%%%%%%%%%
%%%%%%%%%%%%%%%%%%%%%%%%%%%%%%%%%%%%%%%%%%%%%%%%%%%%%%%%%%%%%%%%%%%%%%

\end{document}